\documentclass[12pt]{article}%
\usepackage{graphicx}
\usepackage{amsmath}
\usepackage{amsfonts}
\usepackage{amssymb}
\usepackage{url}%
\usepackage{fullpage}
\begin{document}
\title{Data Driven Computing by the Morphing Fast Fourier Transform Ensemble Kalman
Filter in Epidemic Spread Simulations}
\author{Jan Mandel \and Jonathan D. Beezley \and Loren Cobb \and Ashok Krishnamurthy}
\date{\relax}
\maketitle
\begin{center}\small Department of Mathematical and
Statistical Sciences,\\ University of Colorado Denver, Denver, CO
80217-3364, USA \end{center}
\begin{abstract}
The FFT EnKF data assimilation method is proposed and applied to a stochastic
cell simulation of an epidemic, based on the S-I-R spread model. The FFT EnKF
combines spatial statistics and ensemble filtering methodologies into a
localized and computationally inexpensive version of EnKF with a very small
ensemble, and it is further combined with the morphing EnKF to assimilate
changes in the position of the epidemic.
\end{abstract}
\section{Introduction}

\label{sec:introduction}

Starting a model from initial conditions and then waiting for the result is
rarely satisfactory. The model is generally incorrect, data is burdened with
errors, and new data comes in that needs to be accounted for. This is a
well-known problem in weather forecasting, and techniques to incorporate new
data by sequential statistical estimation are known as data assimilation
\cite{Kalnay-AMD-2003}. The ensemble Kalman filter (EnKF)
\cite{Evensen-2009-DAE} is a popular data assimilation method, which is easy
to implement without any change in the model. The EnKF evolves an ensemble of
simulations, and the model only needs to be capable of exporting its state and
restarting from the state modified by the EnKF. However, the ensemble size
required can be large (easily in the hundreds), the amount of computations in
the EnKF can be significant, special localization techniques need to be
employed to suppress spurious long-range correlations in the ensemble
covariance matrix, and the EnKF does not work well for problems with sharp
coherent features, such as the travelling waves found in epidemics and wildfires.

We propose a variant of EnKF based on the Fast Fourier transform (FFT), which
reduces significantly the amount of computations required by the EnKF, as well
as the ensemble size. The use of FFT is inspired by spatial statistic:
FFT\ EnKF assumes that the state approximately a stationary random field, that
is, the covariance between two points is mainly a function of their distance
vector. Then the multiplication of the covariance matrix and a vector is a
convolution. In addition, the morphing transform \cite{Beezley-2008-MEK} is
used here so that changes of the state both in position and in amplitude are possible.

The FFT\ EnKF with morphing is illustrated here for tracking a simulated
epidemic wave. The use of data assimilation techniques can increase the
accuracy and reliability of epidemic tracking by using the data as soon as
they are available, and some applications of data assimilation in epidemiology
already exist \cite{Bettencourt-2007-TRT,Rhodes-2009-VDA}. The FFT EnKF with
morphing has the potential to reduce complicated simulations and accurate
real-time use of data to a~laptop or a~smartphone in the field.

For FFT EnKF in a~wildfire simulation, see~\cite{Mandel-2010-FFT}. The Fourier
domain Kalman filter (FDKF) \cite{Castronovo-2008-MTC} consists of the Kalman
filter used in each Fourier mode separately.

The covariance of a~stationary random field can be estimated from a single
realization by the covariogram \cite{Cressie-1993-SSD}, which can be computed
efficiently by the FFT \cite{Marcotte-1996-FVC}. We propose to use the
covariogram for an \emph{EnKF with an ensemble of one}, which will be further
developed elsewhere.

\section{FFT EnKF}

\label{sec:fftenkf}

The EnKF approximates the probability distribution of the model state $u$ by
an ensemble of simulations $u_{1},\ldots,u_{N}$. Each member is advanced by
the simulation in time independently. When new data $d$ arrives, it is given
as data likelihood $d\sim N\left(  Hu,R\right)  $, where $H$ is the
\emph{observation operator} and $R$ is the \emph{data error covariance
matrix}. Now the \emph{forecast ensemble} $\left[  u_{k}\right]  $ is combined
with the data by the EnKF analysis \cite{Burgers-1998-ASE}%
\begin{equation}
u_{k}^{a}=u_{k}+C_{N}H^{T}\left(  HC_{N}H^{T}+R\right)  ^{-1}\left(
d+e_{k}-Hu_{k}^{f}\right)  ,\quad k=1,\ldots,N, \label{eq:enkf}%
\end{equation}
to yield the \emph{analysis ensemble} $\left[  u_{k}^{a}\right]  $. Here,
$C_{N}$ is an approximation of the covariance $C$ of the model state, taken to
be the covariance of the ensemble, and $e_{k}$ is sampled from $N\left(
0,R\right)  $. The analysis ensemble is then advanced by the simulations in
time again. In \cite{Mandel-2009-CEK}, it was proved that the ensemble
converges for large $N$ to a sample from the Kalman filtering distribution
when all probability distributions are Gaussian. Of course, the EnKF is used
for more general cases as well.

When $C_{N}$ is the ensemble covariance, the EnKF formulation (\ref{eq:enkf})
does not take advantage of any special structure of the model. This allows a
simple and efficient implementation \cite{Mandel-2009-DAW}, but large
ensembles, often over 100, are needed \cite{Evensen-2009-DAE}. In an
application, variables in the state are \emph{random fields}, and the
covariance decays with spatial distance \cite{Cressie-1993-SSD}.
\emph{Tapering }is the multiplication of sample covariance term-by-term with a
fixed decay function that drops off with the distance. Tapering improves the
accuracy of the approximate covariance for small ensembles
\cite{Furrer-2007-EHP}, but it makes the implementation of (\ref{eq:enkf})
more expensive: the sample covariance matrix can no longer be efficiently
represented as the product of two much smaller dense matrices, but it needs to
be manipulated as a large, albeit sparse, matrix. Random fields in
geostatistics are often assumed to be stationary, that is, the covariance
between two points depends on their spatial distance vector only.

The FFT EnKF discussed here uses a very small ensemble, but larger than one.
We explain the FFT EnKF in the 1D case; higher-dimensional cases are exactly
the same. Consider first the case when the model state consists of one block
only. Denote by $u\left(  x_{i}\right)  $, $i=1,\ldots,n$ the entry of vector
$u$ corresponding to node $x_{i}$. If the random field is stationary, the
covariance matrix satisfies $C\left(  x_{i},x_{j}\right)  =c\left(
x_{i}-x_{j}\right)  $ for some covariance function $c$, and multiplication by
$C$ is the convolution%
\[
v\left(  x_{i}\right)  =\sum_{j=1}^{n}C\left(  x_{i},x_{j}\right)  u\left(
x_{j}\right)  =\sum_{j=1}^{n}u\left(  x_{j}\right)  c\left(  x_{i}%
-x_{j}\right)  ,\quad i=1,\ldots,n.
\]
After FFT, convolution becomes entry-by-entry multiplication of vectors, that
is, multiplication by a diagonal matrix.

We assume that the random field is approximately stationary, so we neglect the
off-diagonal terms of the covariance matrix in the frequency domain, which
leads to the the following FFT EnKF method. First apply FFT to each member,
$\widehat{u}_{k}=Fu_{k}$. Next, approximate the forecast covariance matrix in
the frequency domain by the diagonal matrix with the diagonal entries given by%
\begin{equation}
\widehat{c}_{i}=\frac{1}{N-1}{\sum_{k=1}^{N}}\left\vert \widehat{u}%
_{ik}-\overline{\widehat{u}}_{i}\right\vert ^{2},\quad\text{where\quad
}\overline{\widehat{u}}_{i}=\frac{1}{N}\sum_{k=1}^{N}\widehat{u}_{ik}.
\label{eq:dii}%
\end{equation}
Then define approximate covariance matrix $C_{N}$ by term-by-term
multiplication $\cdot$ in the Fourier domain
\[
u=C_{N}v\iff\widehat{u}=Fu,\quad\widehat{v}=\widehat{c}\bullet\widehat
{u},\quad v=F^{-1}\widehat{v},\quad\left(  \widehat{c}\bullet\widehat
{u}\right)  _{i}=\widehat{c}_{i}\widehat{u}_{i}.
\]
When $H=I$ and $R=rI$, the evaluation of (\ref{eq:enkf}) reduces to%
\begin{equation}
\widehat{u}_{k}^{a}=\widehat{u}_{k}+\widehat{c}\bullet\left(  \widehat
{c}+r\right)  ^{-1}\bullet\left(  \widehat{d}+\widehat{e}_{k}-\widehat{u}%
_{k}^{f}\right)  . \label{eq:spectral-enkf}%
\end{equation}

In general, the state has more than one variable, and $u$, $C$, and $H$ have
the block form%
\begin{equation}
u=\left[
\begin{matrix}
u^{\left(  1\right)  }\\
\vdots\\
u^{\left(  n\right)  }%
\end{matrix}
\right]  ,\quad C=\left[
\begin{matrix}
C^{\left(  11\right)  } & \cdots & C^{\left(  1M\right)  }\\
\vdots & \ddots & \vdots\\
C^{\left(  M1\right)  } & \cdots & C^{\left(  MM\right)  }%
\end{matrix}
\right]  ,\quad H=\left[
\begin{matrix}
H^{\left(  1\right)  } & \cdots & H^{\left(  M\right)  }%
\end{matrix}
\right]  . \label{eq:block-enkf}%
\end{equation}
Here, the first variable is observed, so $H^{\left(  1\right)  }=I$,
$H^{\left(  2\right)  }=0$,\ldots, $H^{\left(  M\right)  }=0$, and
(\ref{eq:enkf}) becomes%
\begin{equation}
u_{k}^{\left(  j\right)  ,a}=u_{k}^{\left(  j\right)  }+C_{N}^{\left(
j1\right)  }\left(  C_{N}^{\left(  11\right)  }+R\right)  ^{-1}\left(
d+e_{k}-u_{k}^{\left(  1\right)  }\right)  ,\quad j=1,\ldots,M,
\label{eq:multi-enkf}%
\end{equation}
and in the frequency domain%
\begin{equation}
\widehat{u}_{k}^{\left(  j\right)  ,a}=\widehat{u}_{k}^{\left(  j\right)
}+\widehat{c}^{\left(  j1\right)  }\bullet\left(  \widehat{c}^{\left(
11\right)  }+r\right)  ^{-1}\bullet\left(  \widehat{d}+\widehat{e}%
_{k}-\widehat{u}_{k}\right)  . \label{eq:spectral-multi-enkf}%
\end{equation}
The cross-covariance between field $j$ and field $1$ is approximated by
neglecting the off-diagonal terms of the sample covariance in the frequency
domain as well,%
\begin{equation}
\widehat{c}_{i}^{\left(  j1\right)  }=\frac{1}{N-1}{\sum_{k=1}^{N}}\left(
\widehat{u}_{ik}^{\left(  j\right)  }-\overline{\widehat{u}}_{i}^{\left(
j\right)  }\right)  \left(  \widehat{u}_{ik}^{\left(  1\right)  }%
-\overline{\widehat{u}}_{i}^{\left(  1\right)  }\right)  ,\quad\text{where}%
\quad\overline{\widehat{u}}_{i}^{\left(  \ell\right)  }=\frac{1}{N}\sum
_{k=1}^{N}\widehat{u}_{ik}^{\left(  \ell\right)  },\ \ell=1,j.
\label{eq:cov-multi-spectral}%
\end{equation}

In the computations reported here, we have used the real sine transform, so
all numbers in (\ref{eq:cov-multi-spectral}) are real. Also, the use of the
sine transform naturally imposes no change of the state on the boundary.

\section{Morphing EnKF}

Given an initial state $u$, the initial ensemble in the morphing
EnKF~\cite{Beezley-2008-MEK,Mandel-2009-DAW} is given by
\begin{equation}
u_{k}^{\left(  i\right)  }=\left(  u_{N+1}^{\left(  i\right)  }+r_{k}^{\left(
i\right)  }\right)  \circ\left(  I+T_{k}\right)  ,\quad k=1,\ldots,N,
\label{eq:initial}%
\end{equation}
with an additional member $u_{N+1}=u$, called the \emph{reference member}. In
(\ref{eq:initial}), $r_{k}^{\left(  i\right)  }$ are random smooth functions
on $\Omega$, $T_{k}$ are random smooth mappings $T_{k}:\Omega\rightarrow
\Omega$, and $\circ$ denotes composition. Thus, the initial ensemble varies
both in amplitude and in position, and the change position is the same in all
blocks. The random smooth functions and mapping are generated by FFT as
Fourier series with random coefficients with zero mean and variance that
decays quickly with frequency.

The data $d$ is an observation of $u^{\left(  1\right)  }$. The first blocks
of all members $u_{1},\ldots,u_{N}$ and $d$ are then registered against the
first block of $u_{N+1}$ as
\[
u_{k}^{\left(  1\right)  }\approx u_{N+1}^{\left(  1\right)  }\circ\left(
I+T_{k}\right)  ,\quad T_{k}\approx0,\quad\nabla T_{k}\approx0,\quad
k=0,\ldots,N,
\]
$u_{0}^{(1)}=d$ and $T_{k}:\Omega\rightarrow\Omega$, $k=0,\ldots,N$ are called
\emph{registration mappings}. The registration mapping is found by multilevel
optimization.
The \emph{morphing transform} maps each ensemble member $u_{k}$ into the
extended state vector, the \emph{morphing representation,}
\begin{equation}
u_{k}\mapsto\widetilde{u}_{k}=M_{u_{N+1}}\left(  u_{k}\right)  =\left(
T_{k},r_{k}^{\left(  1\right)  },\ldots,r_{k}^{\left(  M\right)  }\right)
,\label{eq:morphing-transform}%
\end{equation}
where $r_{k}^{\left(  j\right)  }=u_{k}^{\left(  j\right)  }\circ\left(
I+T_{k}\right)  ^{-1}-u_{N+1}^{\left(  j\right)  }$, $k=0,\ldots,N$, are
\emph{registration residuals}. Likewise, the extended data vector is defined
by $d\mapsto\widetilde{d}=\left(  T_{0},r_{0}^{\left(  1\right)  }\right)  $
and the the observation operator is $\left(  T,r^{\left(  1\right)  }%
,\ldots,r^{\left(  M\right)  }\right)  \mapsto\left(  T,r^{\left(  1\right)
}\right)  $. We then apply the FFT EnKF method (\ref{eq:spectral-multi-enkf})
is applied to the transformed ensemble $\widetilde{u}_{1},\ldots,\widetilde
{u}_{N}$. The covariance $C^{\left(  11\right)  }$ in (\ref{eq:multi-enkf})
consists of three diagonal matrices and we neglect the off-diagonal blocks, so
the fast formula (\ref{eq:spectral-multi-enkf}) can be used. The analysis
ensemble ${u}_{1},\ldots,{u}_{N+1}$ is obtained by the \emph{inverse morphing
transform}%
\begin{equation}
u_{k}^{a,\left(  i\right)  }=M_{u_{N+1}}^{-1}\left(  \widetilde{u}_{k}%
^{a}\right)  =\left(  u_{N+1}^{\left(  i\right)  }+r_{k}^{a,\left(  i\right)
}\right)  \circ\left(  I+T_{k}^{a}\right)  ,\quad k=1,\ldots
,N+1,\label{eq:inverse-morphing}%
\end{equation}
where the new transformed reference member is given by
\begin{equation}
\widetilde{u}_{N+1}^{a}=\frac{1}{N}\sum_{k=1}^{N}\widetilde{u}_{k}%
^{a}.\label{eq:analysis-comp}%
\end{equation}

\section{Epidemic model}

\label{sec:epidemic}

The epidemic model that we used for this study is a spatial version of the
common S-I-R\ dynamic epidemic model. A person is \emph{susceptible}\ or
\emph{infectious}\ in this context if he or she can contract or transmit the
disease, respectively. The \emph{removed}\ state includes those who have
either died, have been quarantined, or have recovered from the disease and
become immune. The state variables are the susceptible ($S$), the infectious
($I$), and the removed ($R$) population densities.The core ideas for
this\ model date back to the 1957 spatial formulation by Bailey
\cite{Bailey-1957-MTE}, but the specific version that we have employed here is
due to Hoppenstaedt \cite[p. 64]{Hoppenstaedt-1975-PDE}.

The population is considered to be dispersed over a planar domain
$\Omega\subset\mathbb{R}^{2}$, and it is labelled according to its position
with respect to the spatial coordinates $x$ and $y$. The (deterministic)
evolution of the state $\left(  S\left(  t\right)  ,I\left(  t\right)
,R\left(  t\right)  \right)  $ is given by%
\begin{equation}
\left.
\begin{array}
[c]{l}%
\frac{\partial S\left(  x,y,t\right)  }{\partial t}=-S\left(  x,y,t\right)
\iint\limits_{\Omega}w\left(  x,y,u,v\right)  I\left(  u,v,t\right)  dudv,\\
\frac{\partial I\left(  x,y,t\right)  }{\partial t}=S\left(  x,y,t\right)
\iint\limits_{\Omega}w\left(  x,y,u,v\right)  I\left(  u,v,t\right)
dudv-q\left(  x,y,t\right)  I\left(  x,y,t\right)  ,\\
\frac{\partial R\left(  x,y,t\right)  }{\partial t}=q_{i}\left(  x,y,t\right)
I\left(  x,y,t\right)  .
\end{array}
\right\}  \label{eq:epi-sir}%
\end{equation}

The function $q\left(  x,y,t\right)  $ gives the rate of removal of infectives
due to death, quarantine, or recovery.\ The weight function $w\left(
x,y,u,v\right)  $ measures the influence of infectives at spatial position
$\left(  u,v\right)  $ on the exposure of susceptibles at position $\left(
x,y\right)  $; in this simulation we used the function $w\left(
x,y,u,v\right)  =\alpha\exp\bigl[-(\left(  x-u\right)  ^{2}+\left(
y-v\right)  ^{2})^{1/2}/\lambda\bigr]$, which expresses the idea that the
influence of nearby infectives decays as an exponential function of Euclidean
distance, with constant $\lambda$, characteristic of the distance at which the
disease spreads. More mobile societies will have larger values of $\lambda$.
The parameter $\alpha$ measures the infectiousness of the disease.

A stochastic cell model is created by treating the quantities on the
right-hand-side of (\ref{eq:epi-sir}) as the intensities of a Poisson process
and by piecewise constant integration over the cells. The domain $\Omega$ is
decomposed into nonoverlapping cells $\Omega_{i}$ with centers $\left(
x_{i},y_{i}\right)  $ and areas $A\left(  \Omega_{i}\right)  $, $i=1,\ldots
,K$. The state in the cell $\Omega_{i}$ is the random element $\left(
S_{i},I_{i},R_{i}\right)  $, advanced in time over the interval $\left[
t,t+\Delta t\right]  $ by%
\[
S_{i}\left(  t+\Delta t\right)  =S_{i}\left(  t\right)  -\Delta S_{i},\quad
I_{i}\left(  t+\Delta t\right)  =I_{i}\left(  t\right)  +\Delta S_{i}-\Delta
R_{i},\quad R_{i}\left(  t+\Delta t\right)  =R_{i}\left(  t\right)  +\Delta
R_{i},
\]
where the random increments $\Delta S_{i}$ and $\Delta R_{i}$ are sampled
from
\begin{align}
\Delta S_{i}  &  \sim\operatorname*{Pois}\left(  S_{i}\left(  t\right)
{\textstyle\sum\nolimits_{j=1}^{K}}
w\left(  x_{i},y_{i},x_{j},y_{j}\right)  I_{j}\left(  t\right)  A\left(
\Omega_{i}\right)  \Delta t\right)  ,\label{eq:delta-Si}\\
\Delta R_{i}  &  \sim\operatorname*{Pois}\left(  q_{i}\left(  t\right)
I_{i}\left(  t\right)  A\left(  \Omega_{i}\right)  \Delta t\right)  ,\nonumber
\end{align}
and $q_{i}\left(  t\right)  $ is the given removal rate in the cell
$\Omega_{i}$. The summation in (\ref{eq:delta-Si}) is done only over the cells
$\Omega_{j}$ near $\Omega_{i}$; for far away cells, the weights $w\left(
x_{i},y_{i},x_{j},y_{j}\right)  $ are negligible. It is not necessary to
compute a Poisson-distributed transmission rate from each source cell to a
given target cell, because a finite sum of independent Poisson-distributed
random variables, each with its own intensity parameter, is itself
Poisson-distributed with an intensity parameter equal to the sum of the
individual intensities.


\begin{figure}[ptb]
\begin{center}%
\begin{tabular}
[c]{cc}%
\includegraphics[width=2.6in]{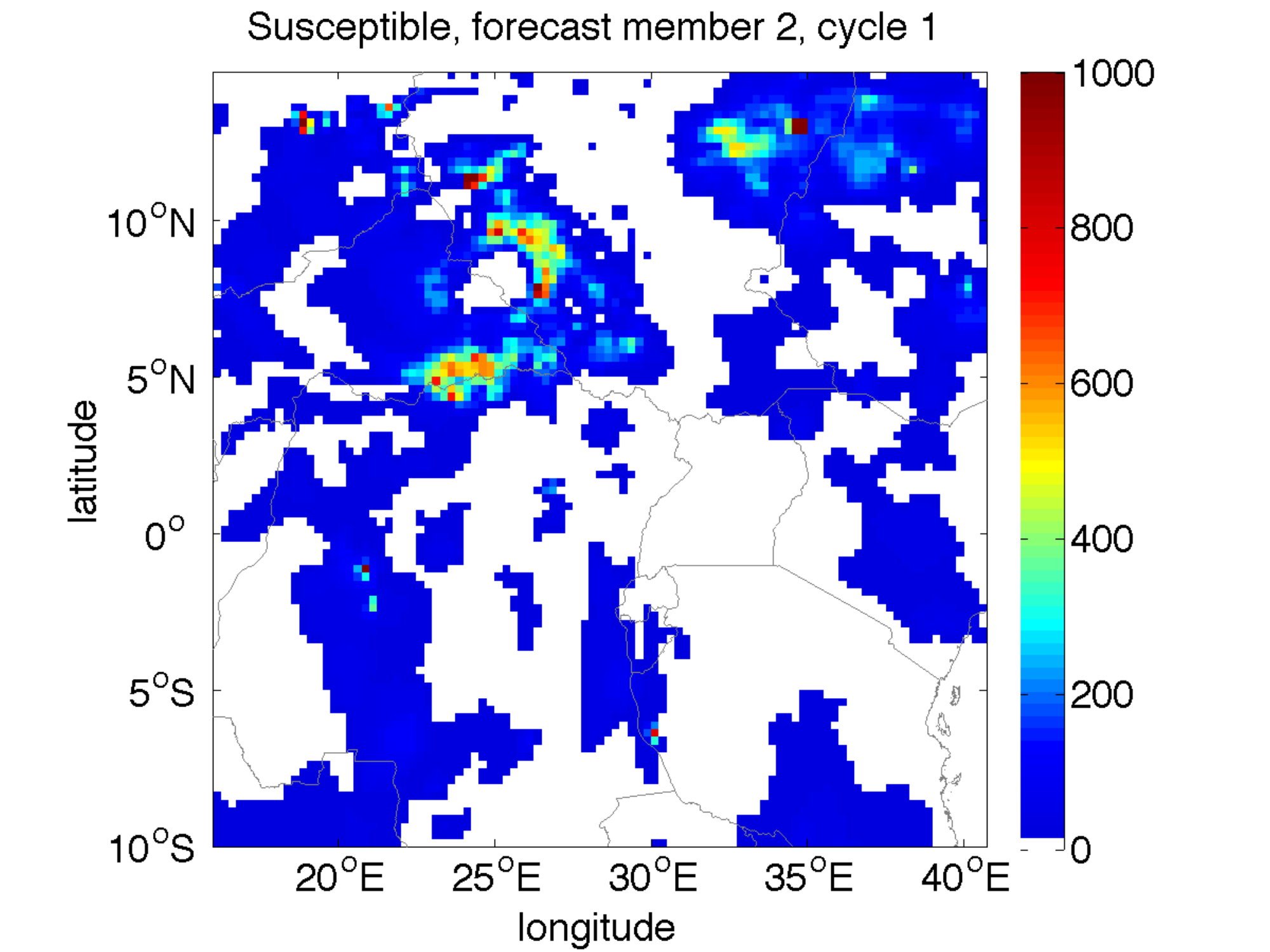} &
\includegraphics[width=2.6in]{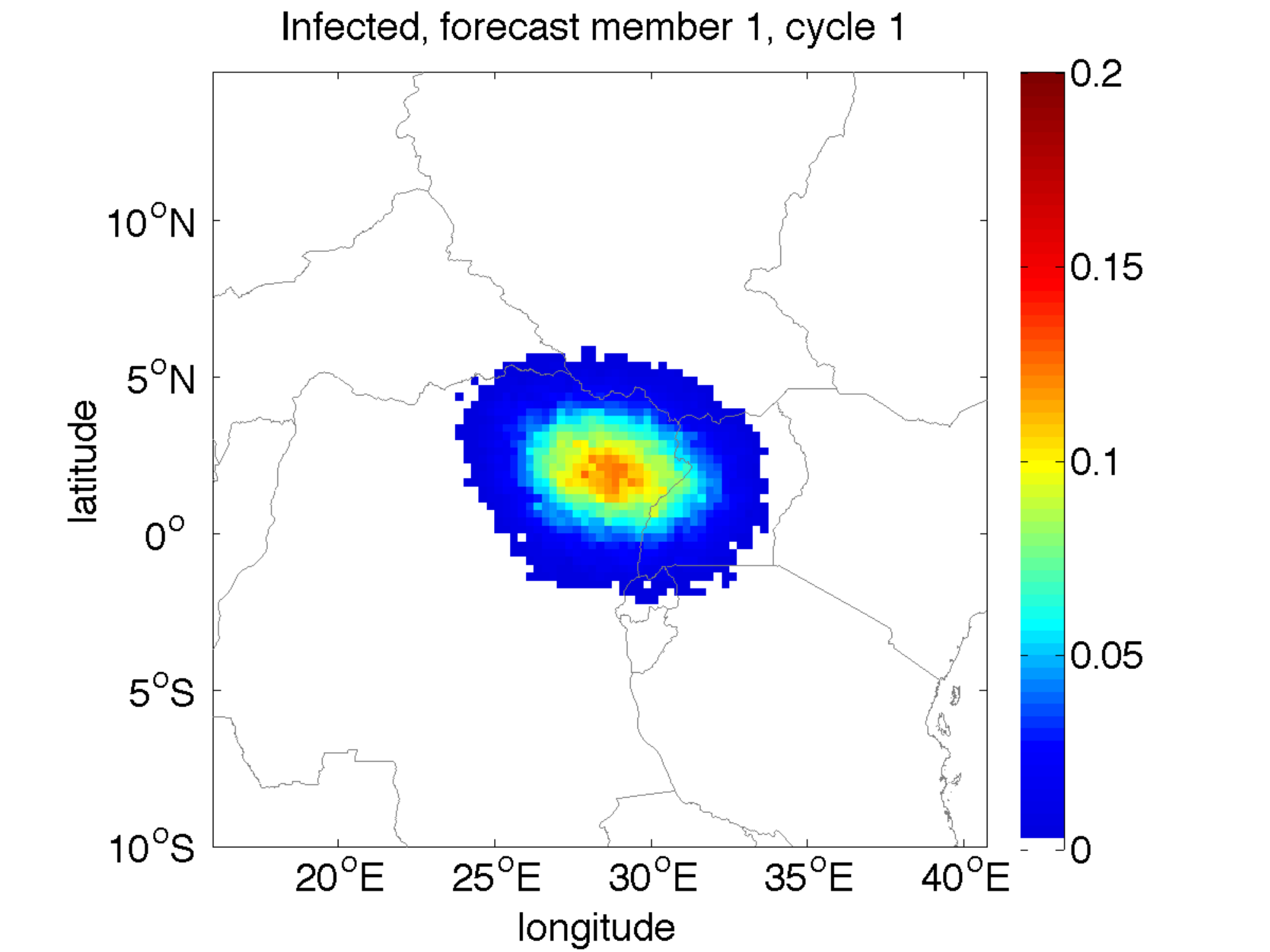}\\
\includegraphics[width=2.6in]{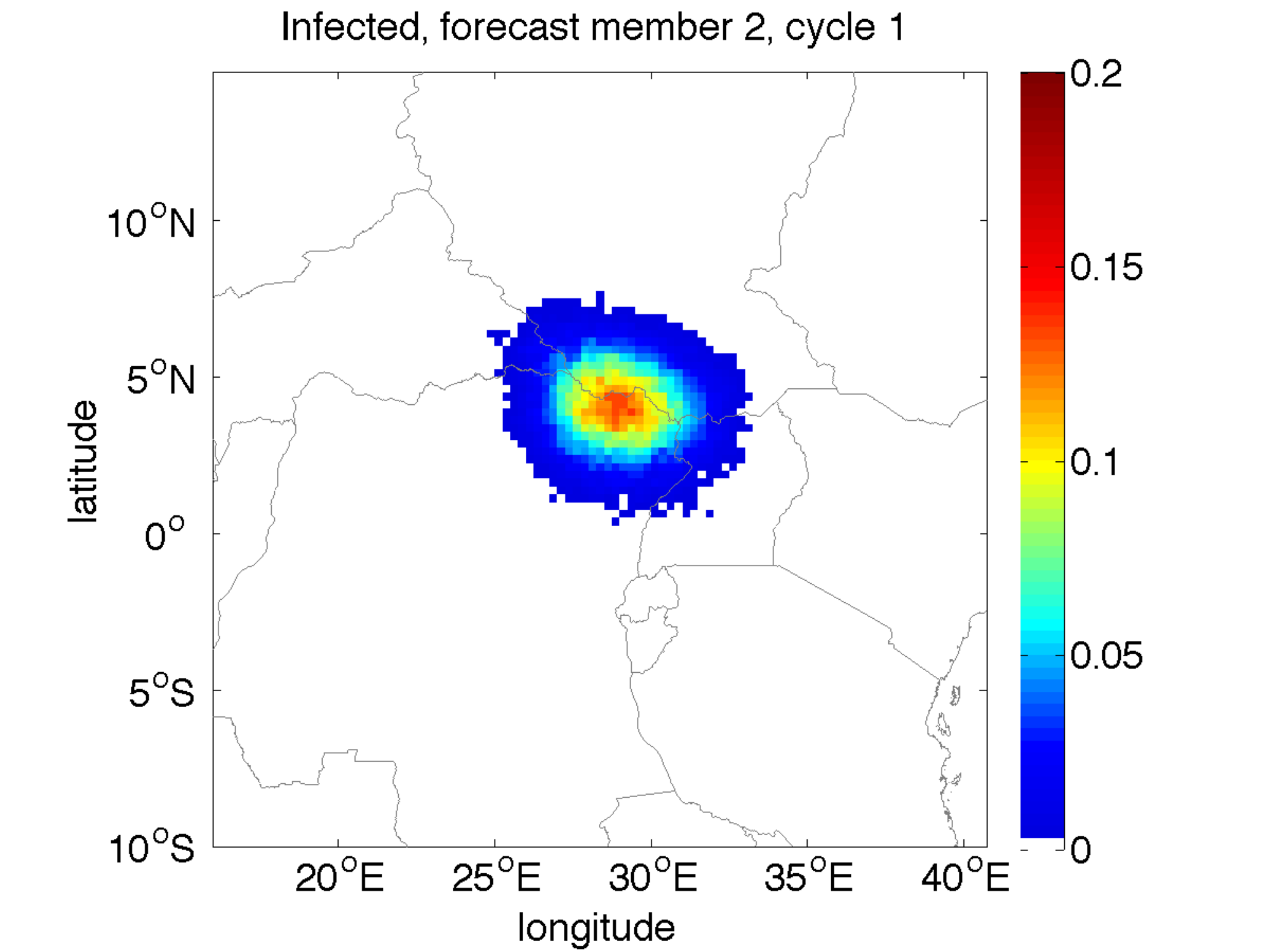} &
\includegraphics[width=2.6in]{figs/fft_morphing_Infected_foremem_1_2.pdf}\\
\includegraphics[width=2.6in]{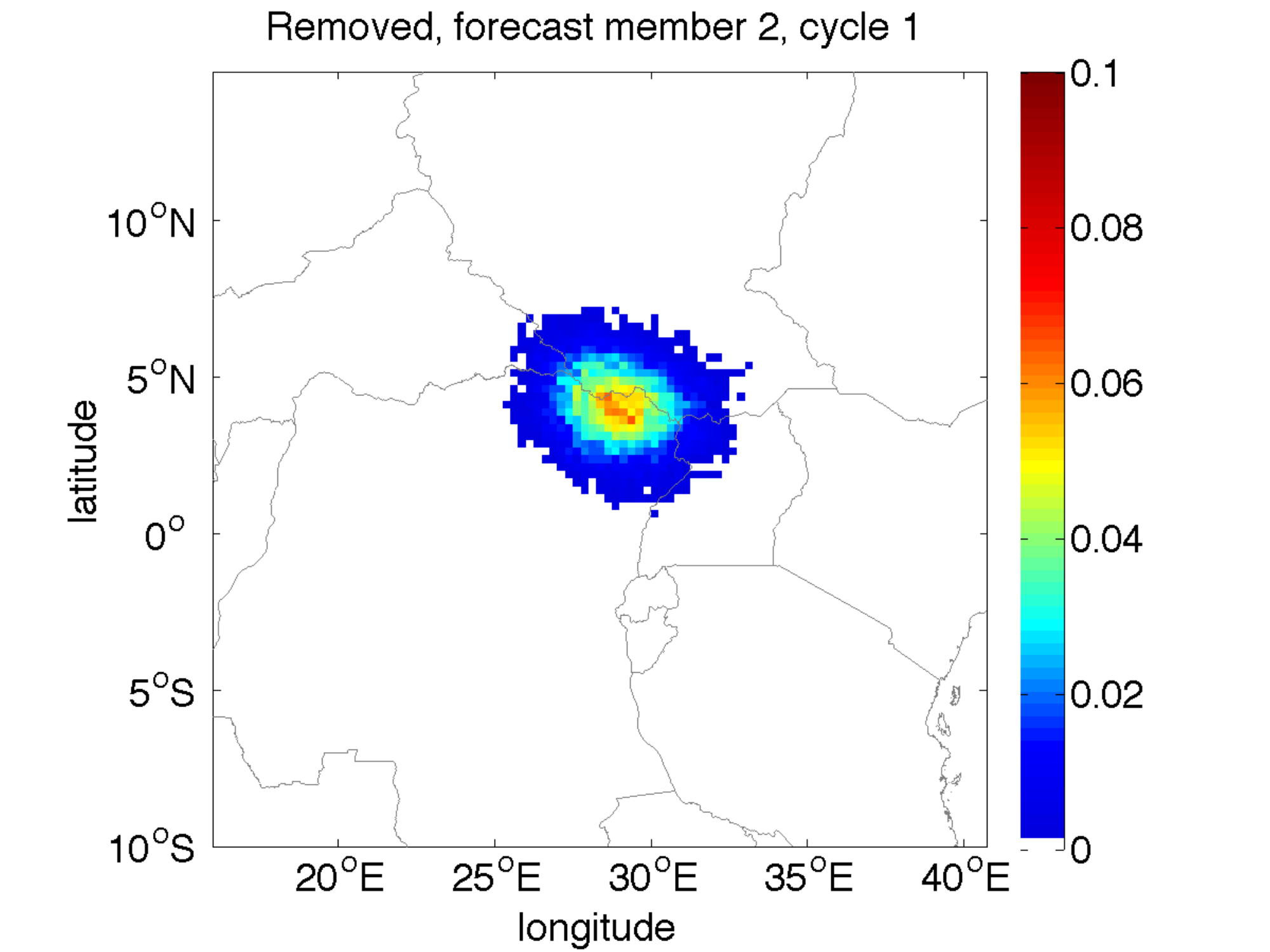} &
\includegraphics[width=2.6in]{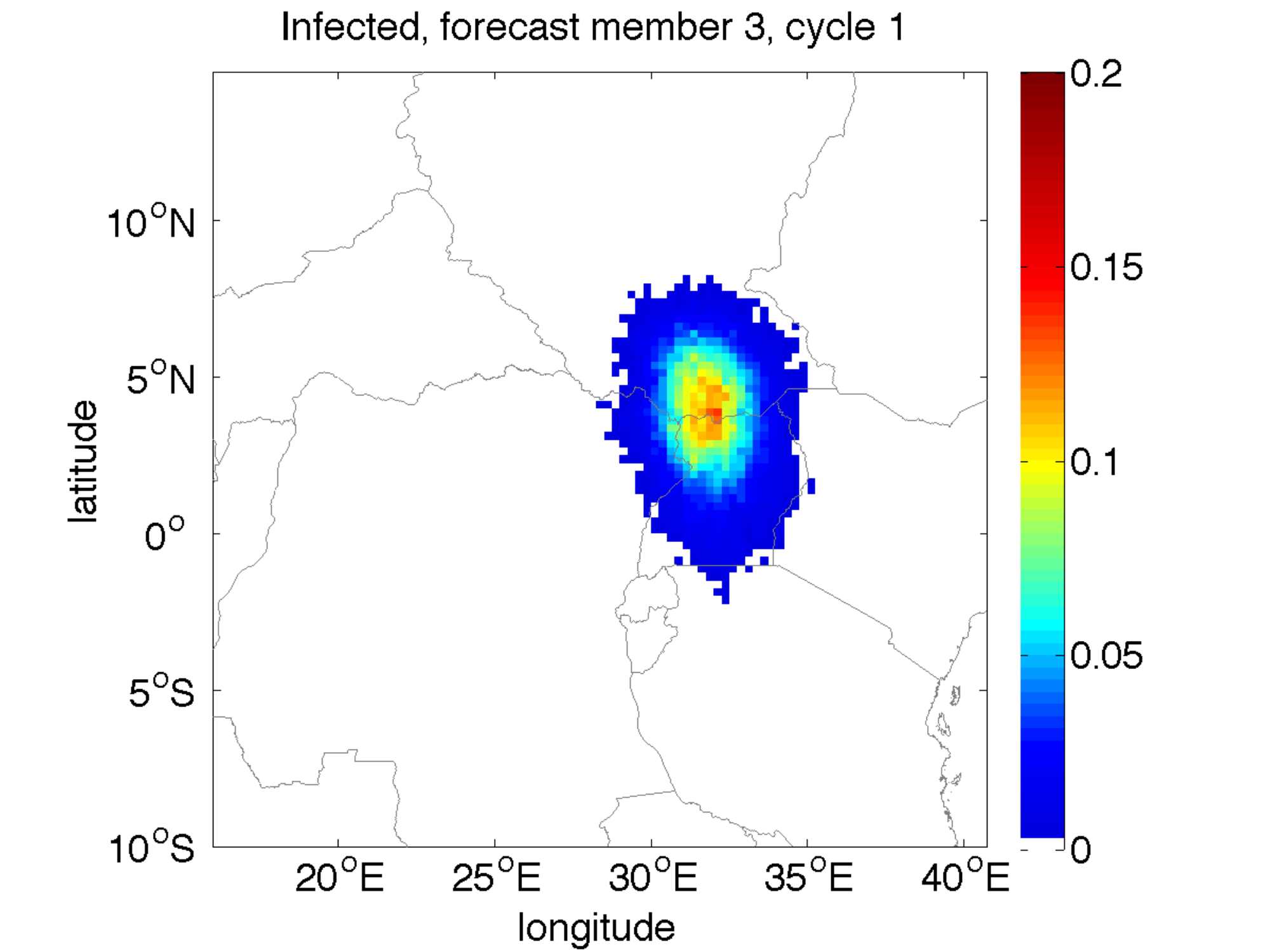}\\
(a) & (b)
\end{tabular}
\end{center}
\caption{(a) The number of people per kilometer squared infected, susceptible,
and removed after $120$ time steps in a simulation of an epidemic disease
spreading through central Africa. These images correspond to variables $I$,
$S$, and $R$ in Equation (\ref{eq:epi-sir}). (b) Number of people infected per
kilometer squared in three forecast ensemble members.}%
\label{fig:epimodel}%
\end{figure}

\section{Computational results}

\label{sec:results}

We have chosen to model an epidemic disease that first emerges in Congo. The
computational domain is a square portion of central Africa. In
Figure~\ref{fig:epimodel} (a), we see the epidemic wave $120$ model time steps
after the emergence of the disease. The behavior of the model is such that any
spurious infection will tend to grow into a secondary infection wave. This is
problematic for data assimilation because the occurrence of spurious features
is virtually guaranteed. We attempt to reduce the occurrence and magnitude of
these features using the morphing transformation and FFT EnKF; however, some
amount of residual artifacts will remain. We have found that by processing the
model state in the following manner, we can further reduce these artifacts. We
begin by scaling the absolute quantities contained in the model variables to a
percentage of the local population before performing the data assimilation.
After data assimilation, we truncate the variables to the range $[0,1]$, and
we apply a threshold so that any infection rate below $1\%$ is set to $0$.
Finally, we rescale the output in absolute units ensuring that the number of
people at each grid cell is preserved. We have applied the FFT EnKF to the
epidemic model described in Section \ref{sec:epidemic} with an ensemble of
size $5$. Each ensemble simulation was started with the same initial
conditions, but with different random seeds, and advanced in time by $100$
model time units, then perturbed randomly to obtain the initial ensemble. The
analysis ensemble and data were advanced in time an additional 20 model time
steps for further assimilation cycles. In total, 3 assimilation cycles were
performed in this manner.

We have perturbed each member of the initial ensemble randomly in space by
applying (\ref{eq:inverse-morphing}) to the each variable of the morphing
representation of the model. The mappings $T_{k}$ for this perturbation were
generated from a space of smooth functions that are zero at the boundary.
While the residuals $r_{k}$ are customarily initialized to smooth random
fields as well, we have chosen to set $r_{k}=0$ to avoid spurious infections.
We instead multiply each field after the inverse morphing transform by
$1+s_{k}$, where $s_{k}$ is another smooth random field. This ensures that an
initial infection rate of $0$ is unchanged by the perturbation. A part of a
typical ensemble with spatial as well as amplitude variability is shown
Figure~\ref{fig:epimodel} (b).

The output of the observation function used in this example consists of the
\emph{Infected} field of the model. In this case, the data is a spatial
\textquotedblleft image\textquotedblright\ of the number of infected persons
in each grid cell. The data were generated synthetically from a model
simulation, which was initialized in the same manner as the ensemble.

\begin{figure}[ptb]
\begin{center}%
\begin{tabular}
[c]{cc}%
\includegraphics[width=2.6in]{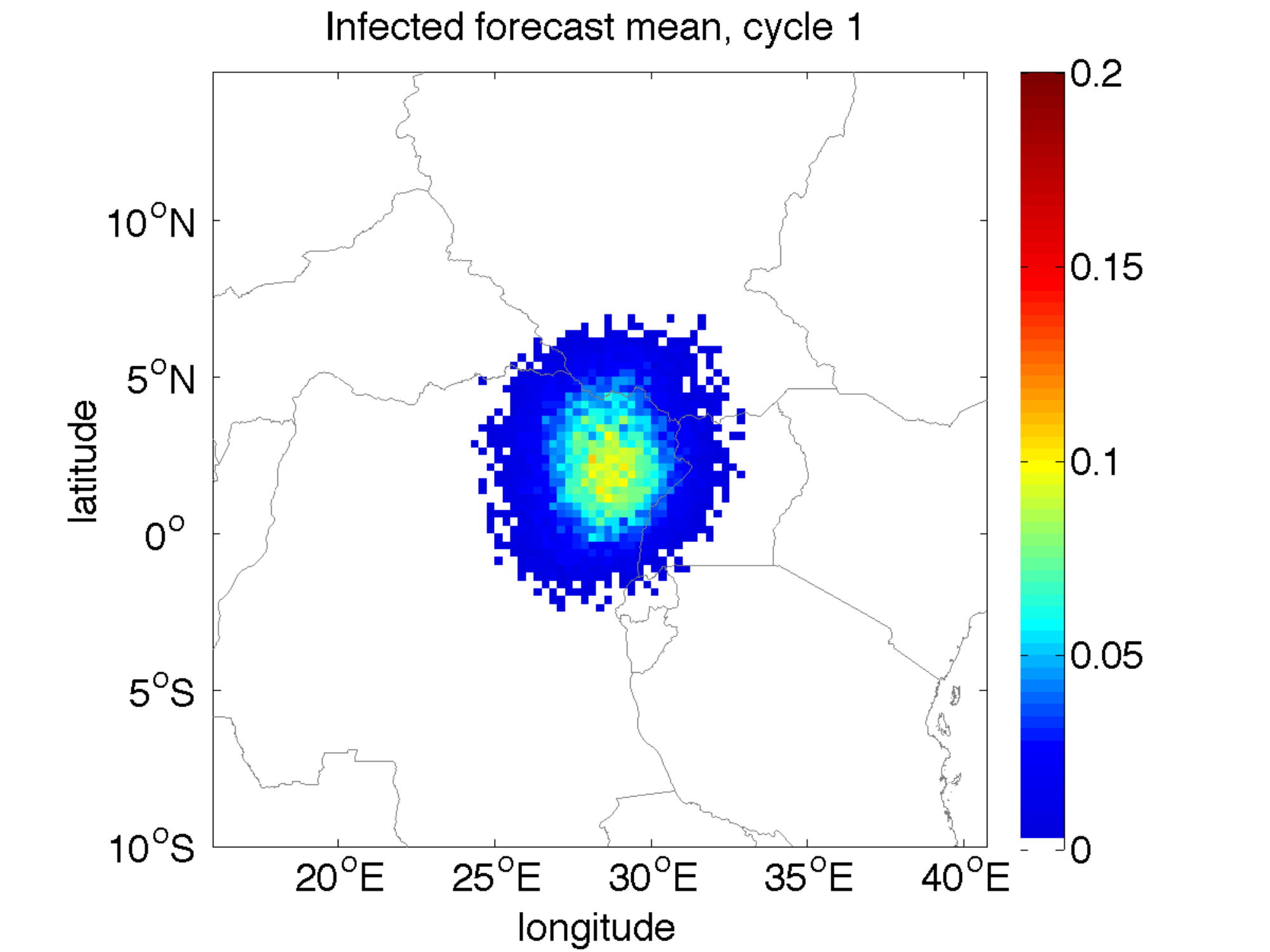} &
\includegraphics[width=2.6in]{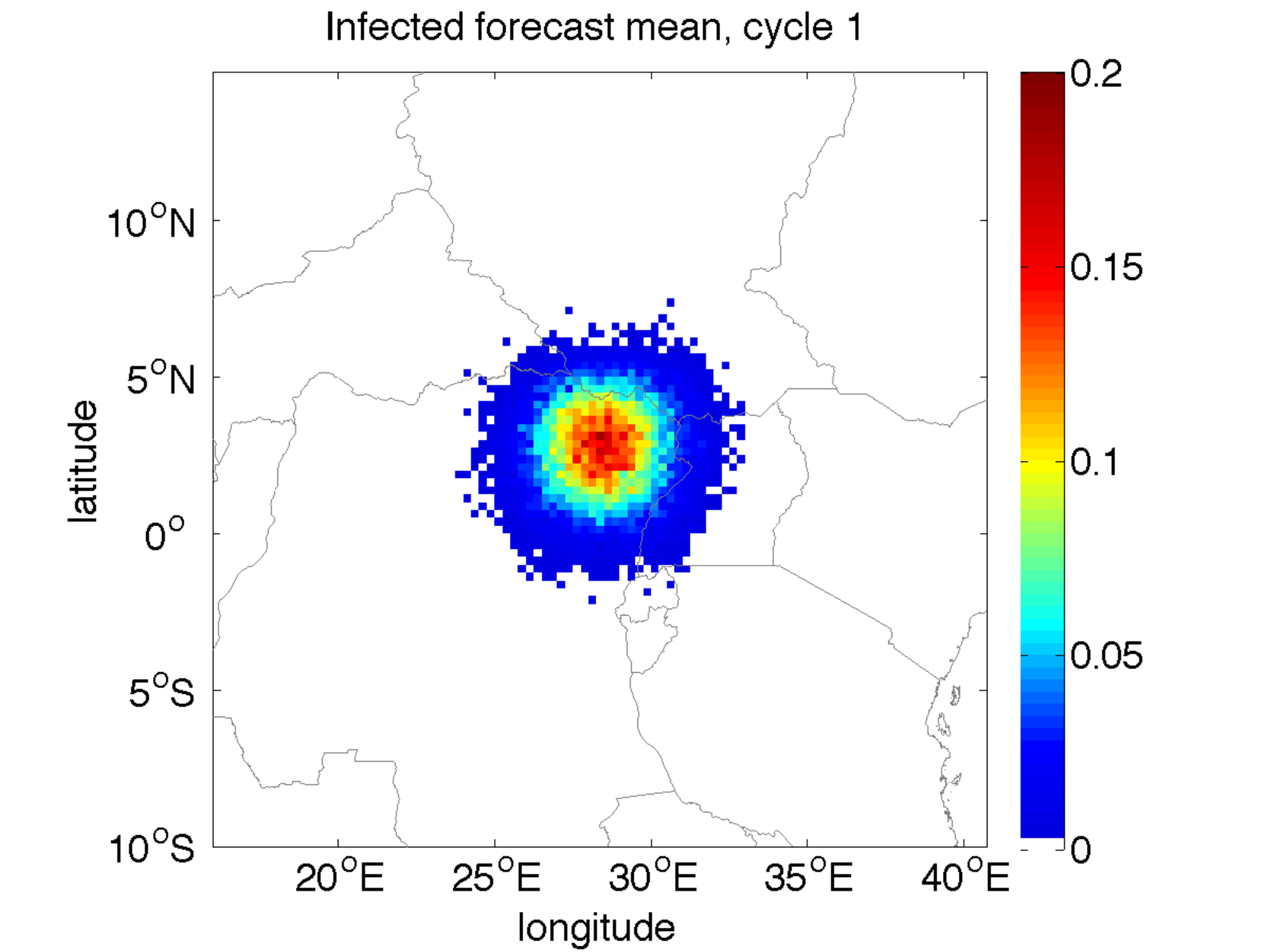}\\
\includegraphics[width=2.6in]{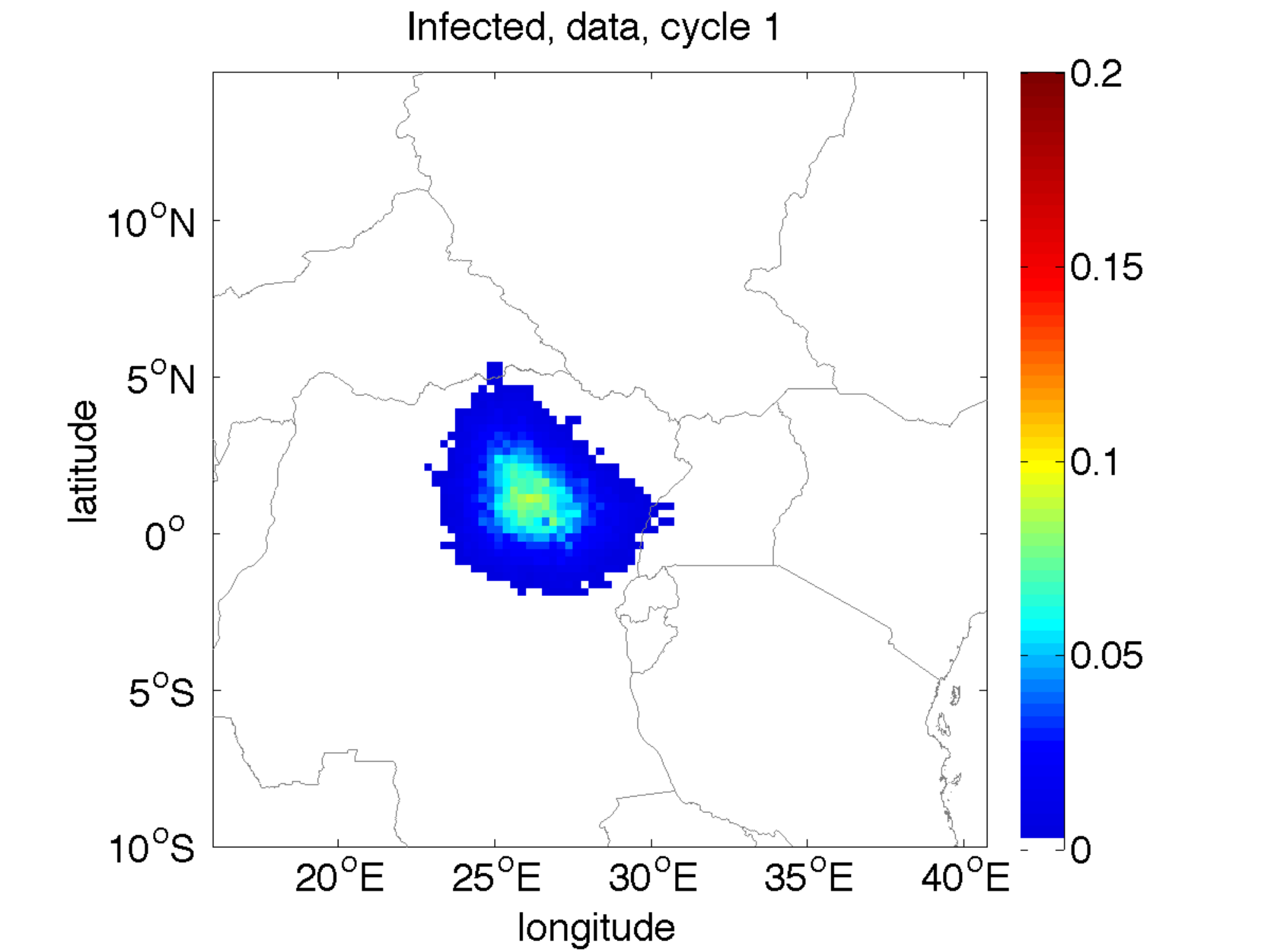} &
\includegraphics[width=2.6in]{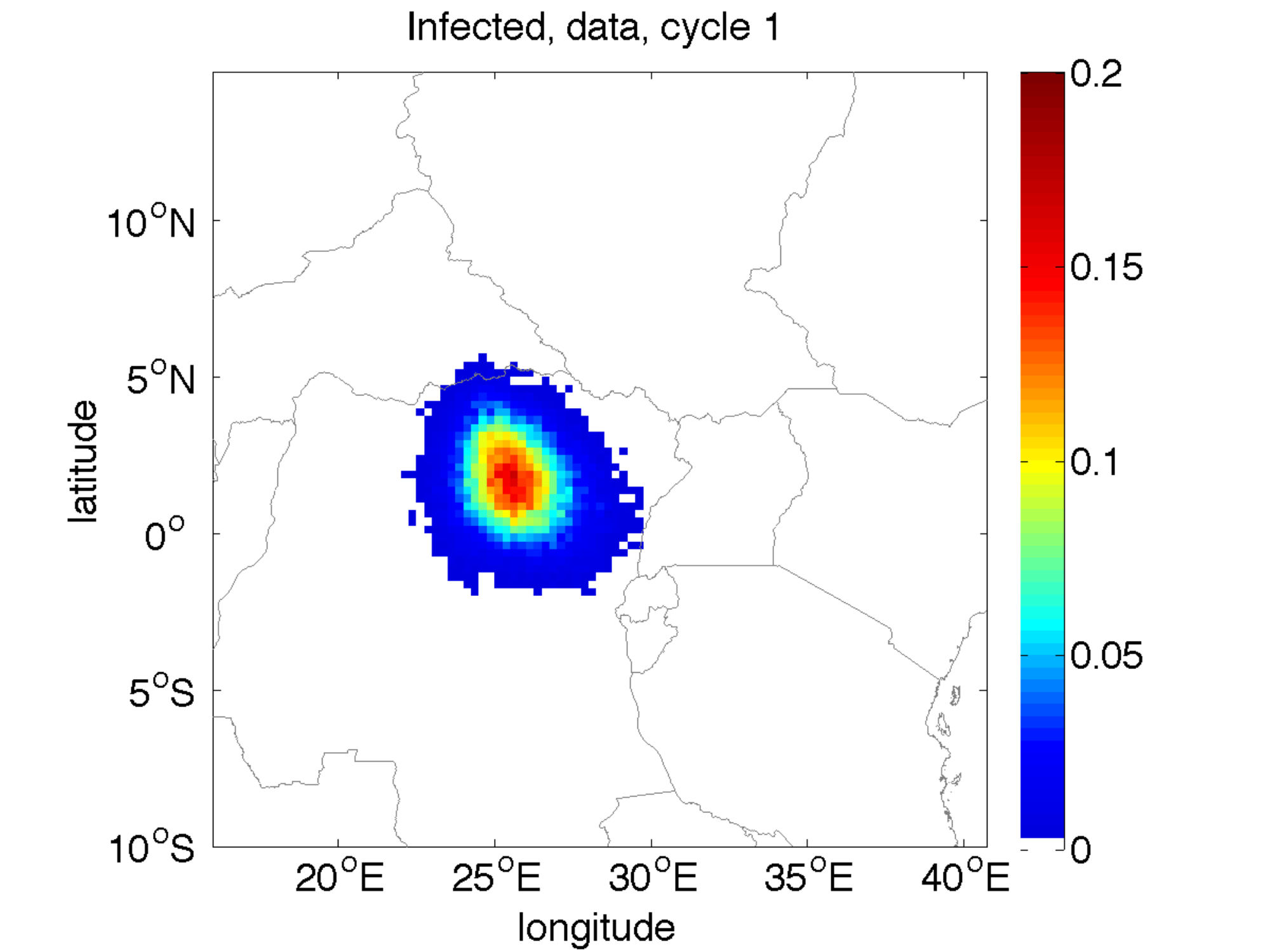}\\
\includegraphics[width=2.6in]{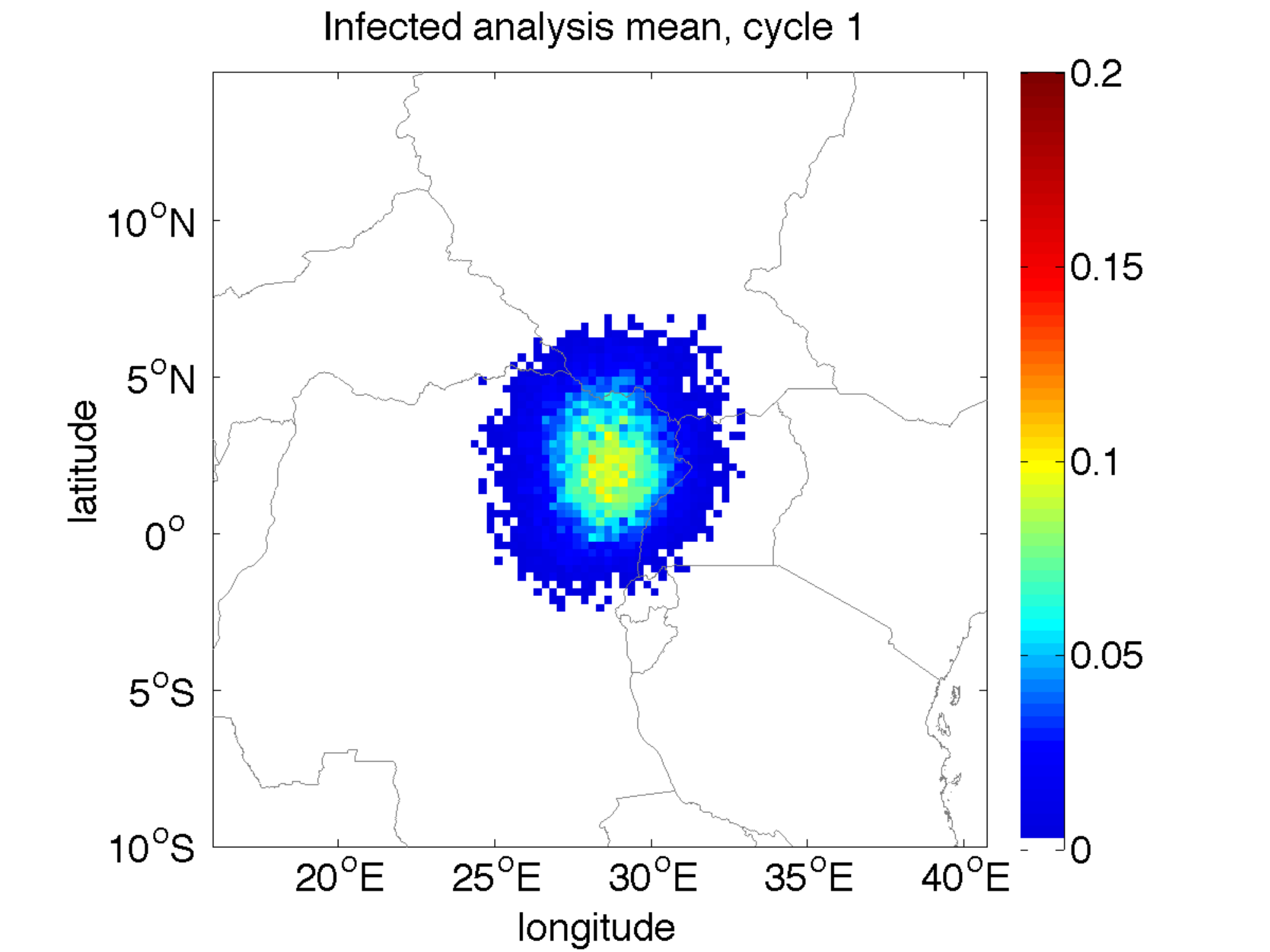} &
\includegraphics[width=2.6in]{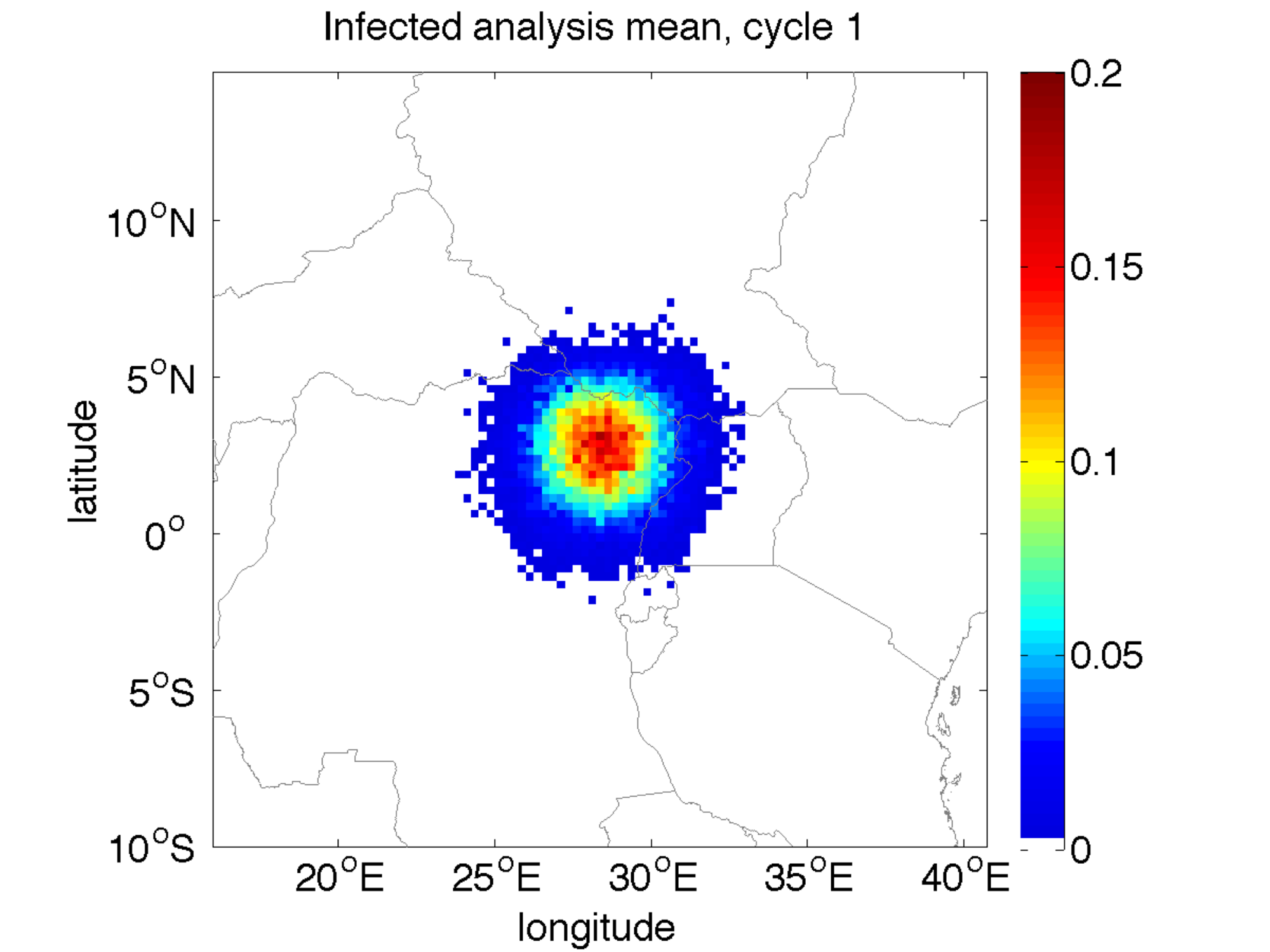}\\
(a) & (b)
\end{tabular}
\end{center}
\caption{The number of people infected per kilometer squared in analysis cycle
1 using the standard EnKF and FFT EnKF, each with ensemble size of $5$. Both
approaches are unable to move the location of the infection in the simulation
state.}%
\label{fig:nomorphing}%
\end{figure}

\begin{figure}[ptb]
\begin{center}%
\begin{tabular}
[c]{cc}%
\includegraphics[width=2.6in]{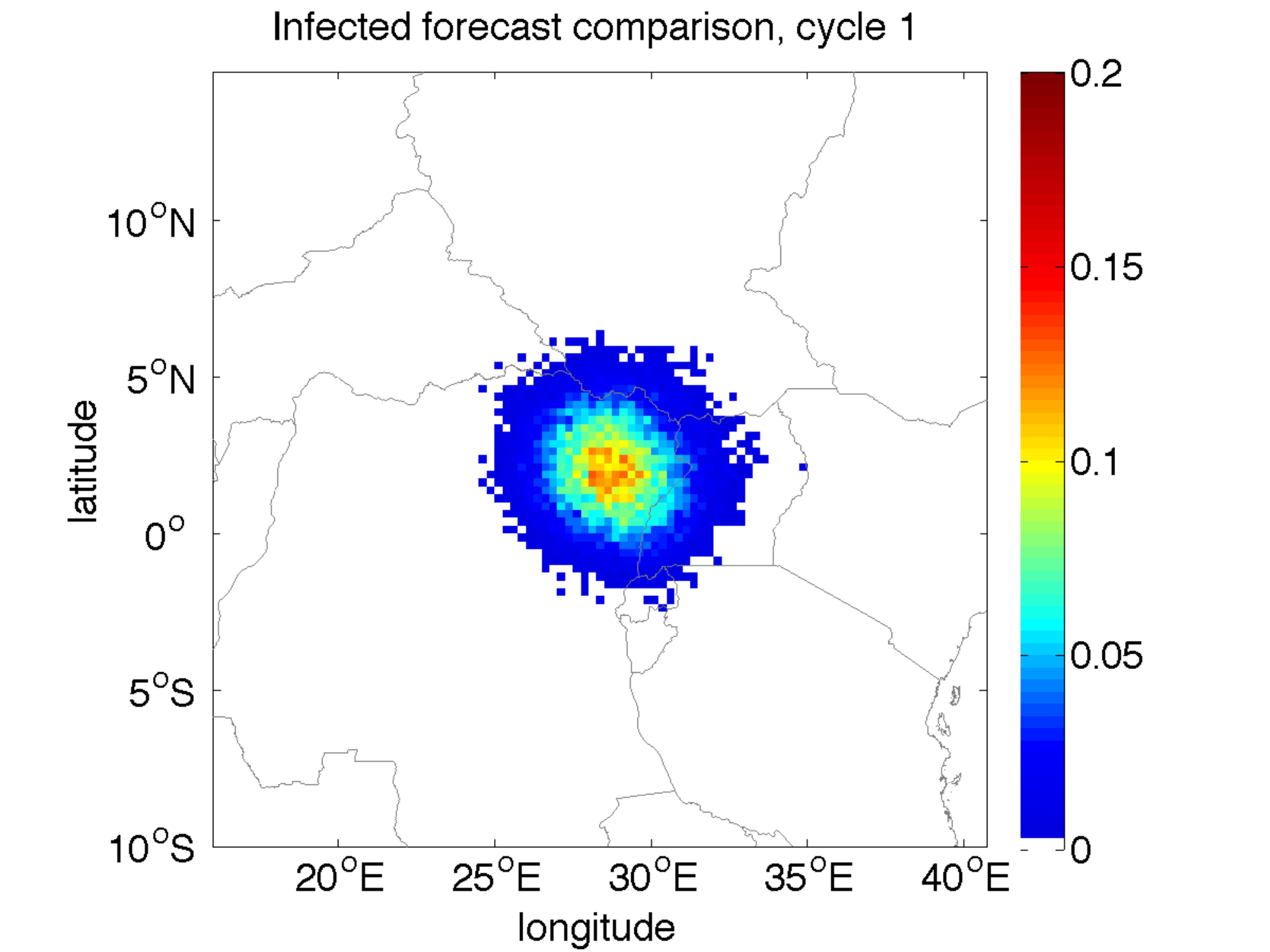} &
\includegraphics[width=2.6in]{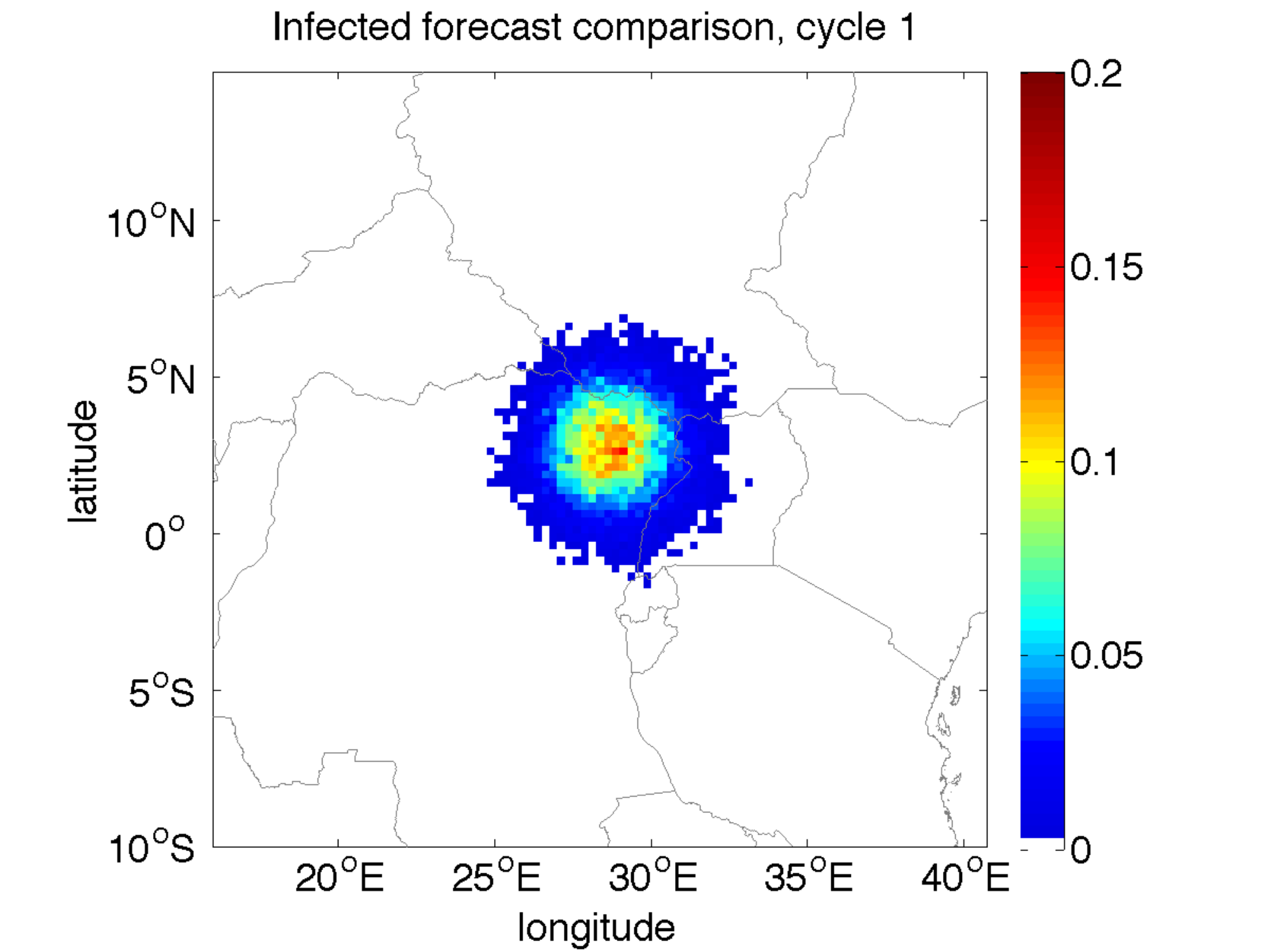}\\
\includegraphics[width=2.6in]{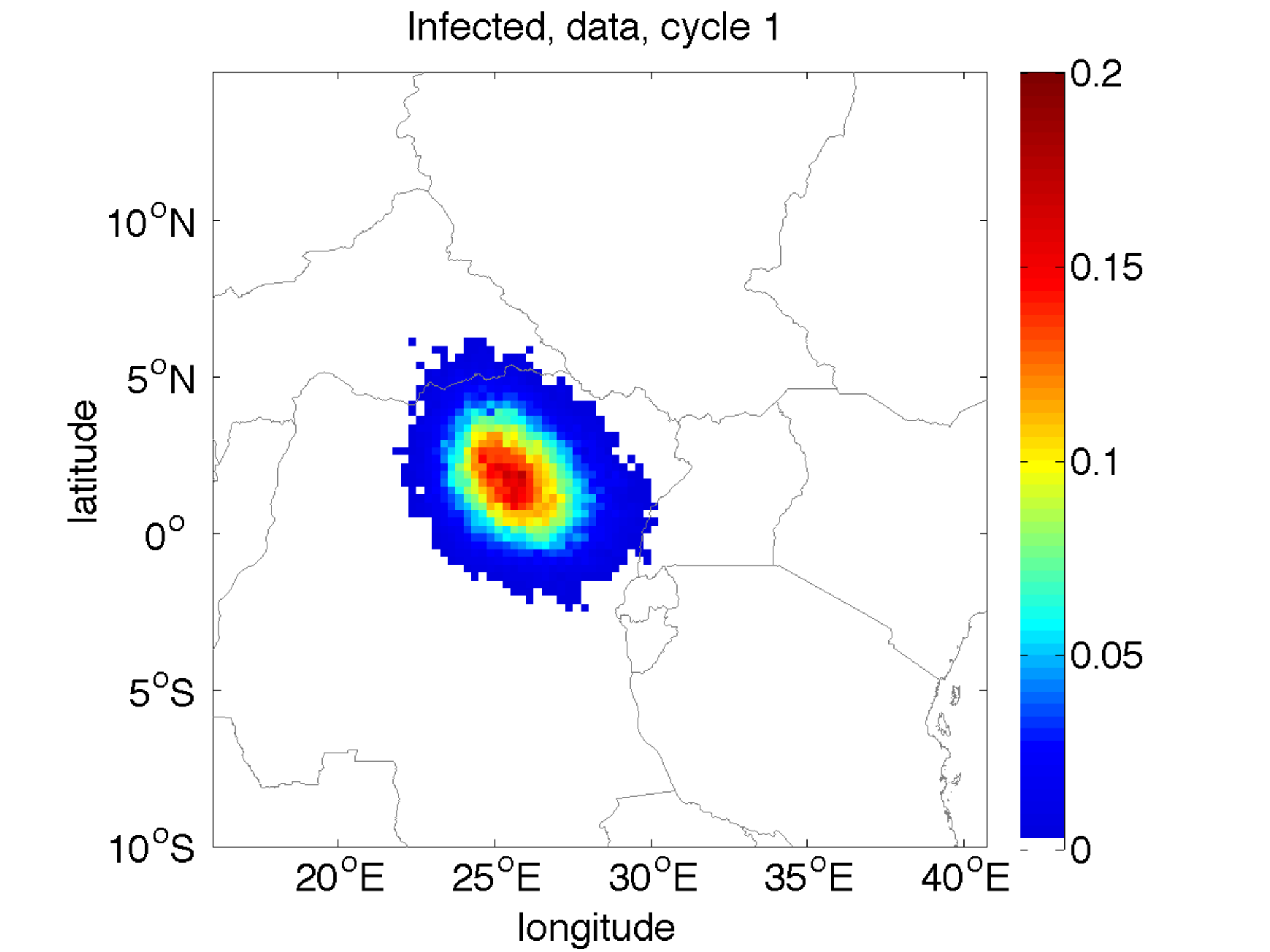} &
\includegraphics[width=2.6in]{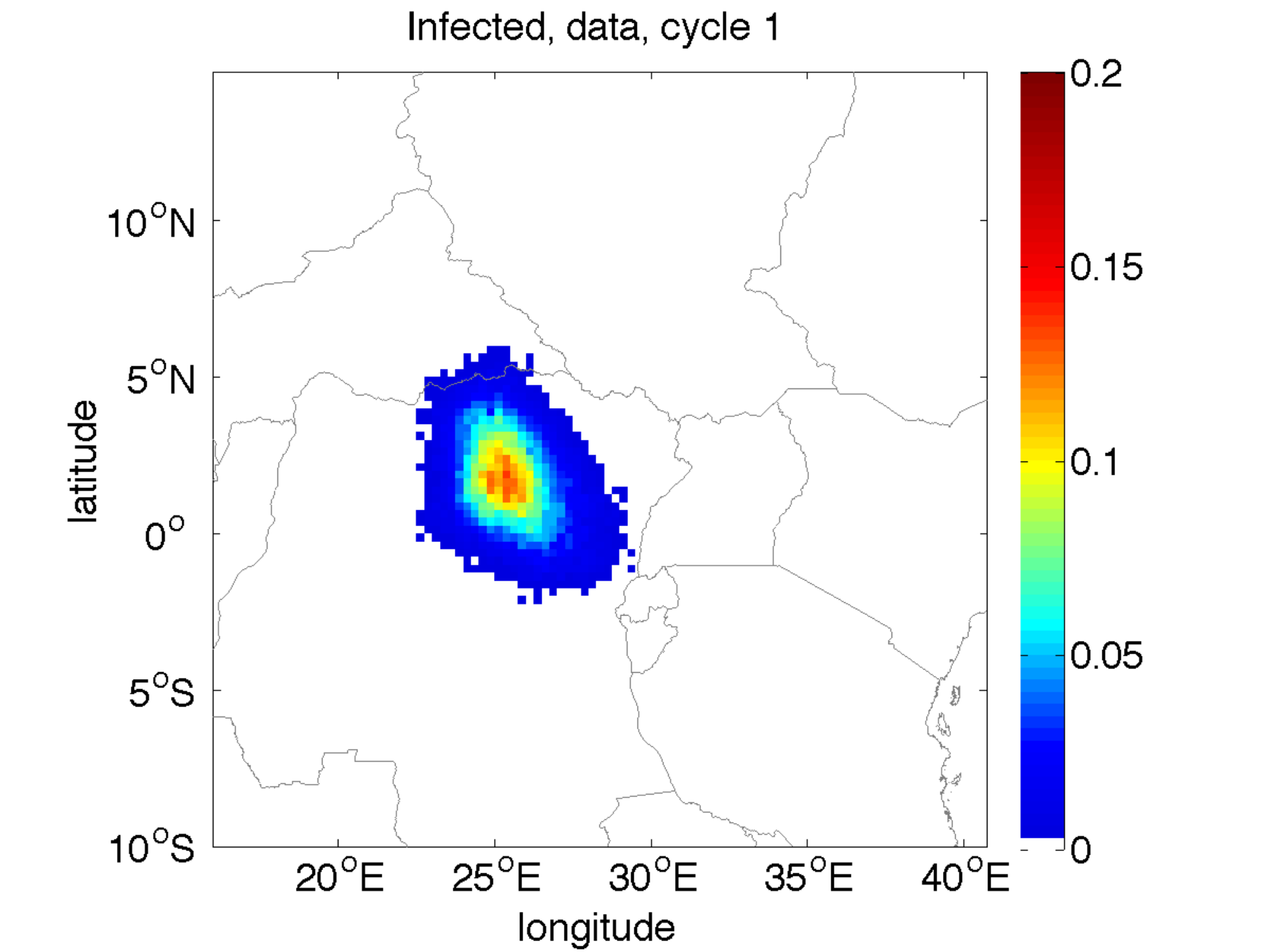}\\
\includegraphics[width=2.6in]{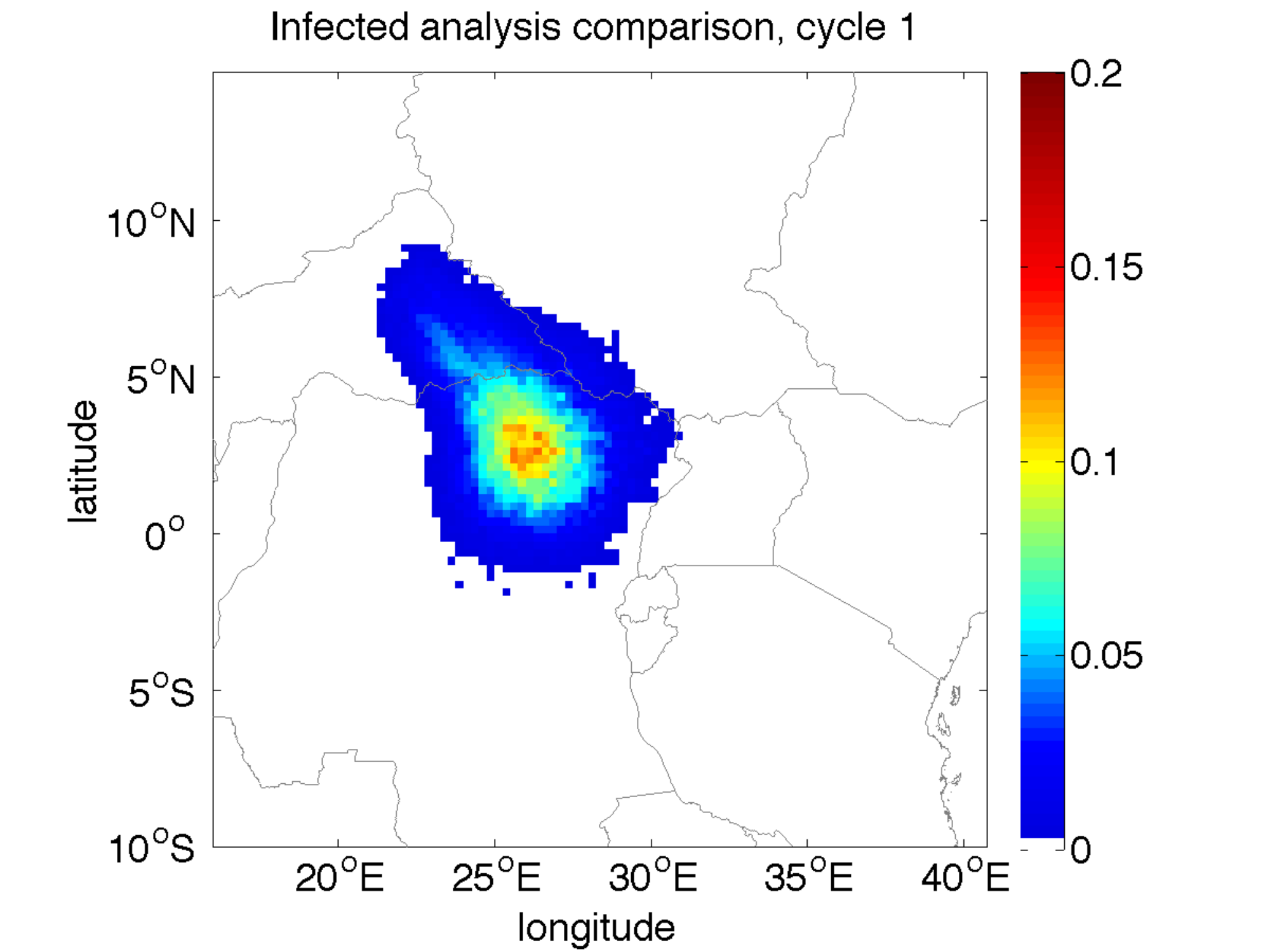} &
\includegraphics[width=2.6in]{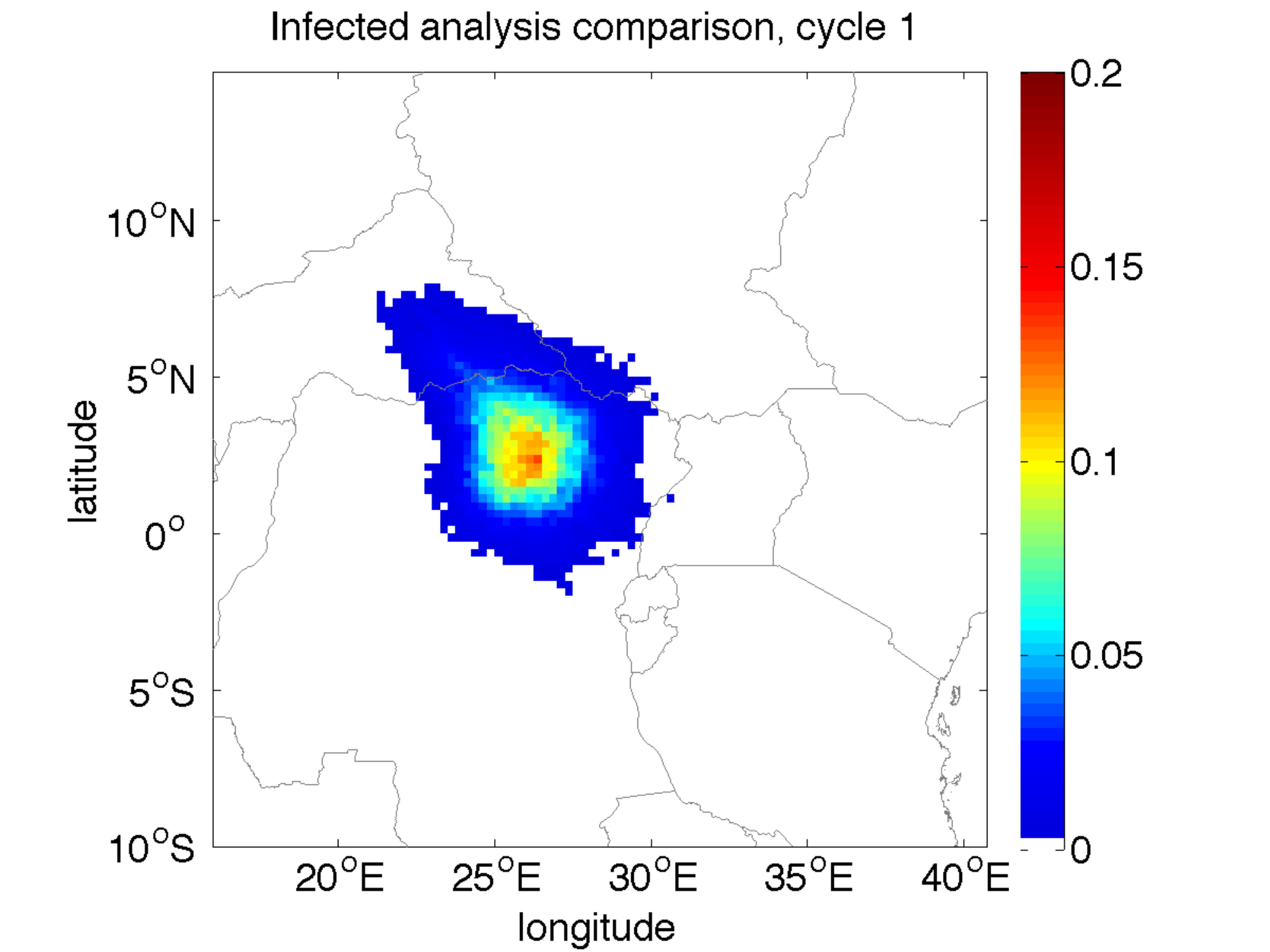}\\
(a) & (b)
\end{tabular}
\end{center}
\caption{The number of people infected per kilometer squared in analysis cycle
1 using the morphing EnKF and morphing FFT EnKF, each with ensemble size of
$5$. Both approaches are able to move the state spatially and perform
similarly. However, EnKF suffers from stronger artifacts due to low accuracy
and low rank of the ensemble covariance than the morphing FFT EnKF. }%
\label{fig:withmorphing}%
\end{figure}

Four variants of the EnKF were then applied: the standard EnKF and FFT EnKF
and morphing EnKF and morphing FFT EnKF. The same initial ensemble and the
same data were used for each method. The deviation of the initial ensemble and
the model error were chosen so that the analysis should be about half way
between the forecast and the data. In the morphing variants, the data
deviation in the amplitude was taken very large, so that the filter updates
essentially only the position. Ensemble of size $5$ was used. The result in
the first assimilation cycle for each method is shown in
Figures~\ref{fig:nomorphing} and \ref{fig:withmorphing}. The first image in
each column is the forecast mean. In the morphing variants, the mean is taken
over all ensemble members in all fields of the morphing representation
(\ref{eq:morphing-transform}) and it plays the role of the comparison state
for registration. Thus, in the morphing variants, both the amplitude and the
position of the infection wave in the ensemble members are averaged. The
second image in each column is the data, which is a model trajectory started
from the same initial state for each method. Because the model is itself
stochastic, the data images are slightly different. The third image in each
column is the analysis mean, which is taken in the morphing representation
(\ref{eq:analysis-comp}) for two morphing filters, so that both the amplitude
and the location are averaged.

We see that both standard EnKF and FFT\ EnKF filters cannot move the state
towards the data; a much larger ensemble would be needed. The morphing EnKF
does move the state towards the data, but there are strong artifacts due to
the poor approximation of the covariance by the covariance of the small
ensemble. Finally, the morphing FFT-EnKF is capable of moving the state
towards the data better.

\qquad

\section{Conclusion}

We have introduced morphing FFT EnKF and presented preliminary results on data
assimilation for an epidemic simulation. Morphing was essential to move the
state towards the data, but it resulted in artifacts for the small ensemble
size used, yet small ensemble size is important to perform simulations with
data assimilation on general computing devices instead of supercomputers. We
have observed that the estimation of the covariance matrix in the frequency
domain results in better forecast covariance in the algorithm, which has the
potential to reduce the artifacts due to small ensemble size.

\pagebreak

\section{Acknowledgements}

This work was partially supported by NIH grant 1 RC1 LM01641-01 and NSF grants
CNS-0719641 and ATM-0835579.


\bibliographystyle{elsarticle-num}
\bibliography{fftenkf-iccs2010,../../bibliography/dddas-jm}

\end{document}